\pdfoutput=1

\RequirePackage{amsthm}

\documentclass[sn-mathphys-num]{sn-jnl}%

\usepackage{pifont}%
\usepackage{graphicx}%
\usepackage{multirow}%
\usepackage{amsmath,amssymb,amsfonts}%
\usepackage{amsthm}%
\usepackage{mathrsfs}%
\usepackage[title]{appendix}%
\usepackage{xcolor}%
\usepackage{textcomp}%
\usepackage{manyfoot}%
\usepackage{booktabs}%
\usepackage{algorithm}%
\usepackage{algorithmicx}%
\usepackage{algpseudocode}%
\usepackage{listings}%
\usepackage{caption}%
\usepackage{mathtools}
\usepackage{microtype}
\usepackage[T1]{fontenc}
\DeclareMathOperator{\DFT}{DFT}
\DeclareMathOperator{\PAF}{PAF}
\DeclareMathOperator{\PSD}{PSD}
\DeclareMathOperator{\PG}{PG}
\newcommand{\circledequal}{\raisebox{0.18ex}{\footnotesize$(=)$}}

\theoremstyle{thmstyleone}%
\newtheorem{theorem}{Theorem}%

\newtheorem{lemma}{Lemma}

\newtheorem{conjecture}{Conjecture}

\newtheorem{example}{Example}%
\newtheorem{corollary}{Corollary}%

\theoremstyle{thmstylefour}
\newtheorem*{proofalt}{Proof}

\renewcommand{\i}{\mathbf{i}}

\newcommand{\com}{/\hspace{-1pt}/ }

\DeclareMathOperator{\Match}{\textsc{Match}}
\DeclareMathOperator{\Uncompress}{\textsc{Uncompress}}
\DeclareMathOperator{\Equiv}{\textsc{Equiv}}
\DeclareMathOperator{\Filter}{\textsc{Filter}}

\gdef\orcidlogo{\includegraphics[width=3mm]{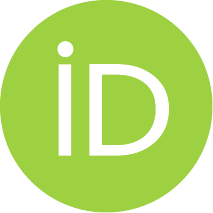}}
\renewcommand{\orcid}[1]{\href{https://orcid.org/#1}{\orcidlogo}}

\newcommand\blfootnote[1]{%
  \begingroup
  \renewcommand\thefootnote{}\footnote{#1}%
  \addtocounter{footnote}{-1}%
  \endgroup
}

\begin{document}

\title{New Results on Periodic Golay Pairs}

\author[1]{\fnm{Tyler} \sur{Lumsden}}\email{lumsdent@uwindsor.ca}

\author[2]{\fnm{Ilias} \sur{Kotsireas}\orcid{0000-0003-2126-8383}}\email{ikotsireas@wlu.ca}

\author[1]{\fnm{Curtis} \sur{Bright}\orcid{0000-0002-0462-625X}}\email{cbright@uwindsor.ca}

\affil[1]{\orgdiv{School of Computer Science}, \orgname{University of Windsor}, \country{Canada}}

\affil[2]{\orgdiv{CARGO Lab}, \orgname{Wilfrid Laurier University}, \country{Canada}}

\abstract{In this paper, we provide algorithmic methods for conducting exhaustive searches for periodic Golay pairs. Our methods enumerate several lengths beyond the currently known state-of-the-art available searches: we conducted exhaustive searches for periodic Golay pairs of all lengths $v \leq 72$ using our methods, while only lengths $v \leq 34$ had previously been exhaustively enumerated.
Our methods are applicable to periodic complementary sequences in general.
We utilize sequence compression, a method of sequence generation derived in 2013 by Đoković and Kotsireas.
We also introduce and implement a new method of ``multi-level'' compression, where sequences are uncompressed in several steps.
This method allowed us to exhaustively search all lengths $v \leq 72$ using less than 10 CPU years.
For cases of complementary sequences where uncompression is not possible,
we introduce some new methods of sequence generation inspired by the isomorph-free exhaustive generation algorithm of orderly generation.
Finally, we pose a conjecture regarding the structure of periodic Golay pairs and prove it holds in many lengths, including all lengths $v<100$.
We demonstrate the usefulness of our algorithms by providing the first ever examples of periodic Golay pairs of length $v = 90$.
The smallest length for which the existence of periodic Golay pairs is undecided is now $106$.}

\keywords{Periodic Golay pairs, Compression, Orderly generation, Complementary sequences, Algorithms, Computer Science}

\pacs[MSC Classification]{11B83, 05B20, 94A55, 68W30}

\maketitle

\blfootnote{Curtis Bright was supported by an NSERC Discovery Grant and Tyler Lumsden was supported by an NSERC undergraduate research assistantship.}

\section{Introduction}\label{sec:intro}
Introduced by Marcel Golay in 1949~\cite{Golay1949}, Golay pairs are used for their relevance in various communications applications, such as multi-slit spectrometry.
Golay pairs consist of two $\{-1,+1\}$-sequences $(A, B)$, each of length~$v$, such that their respective aperiodic autocorrelations sum to zero for all nontrivial shifts.
Periodic Golay pairs are a periodic analog of Golay pairs.
In particular, periodic Golay pairs consist of two $\{-1,+1\}$-sequences $(A, B)$, each of length~$v$, such that their respective \emph{periodic} autocorrelations sum to zero
for all nontrivial shifts. 

Periodic Golay pairs are also of independent mathematical interest; for example,
periodic Golay pairs of length~$v$ can be used to construct Hadamard matrices of order~$2v$~\cite{NA:2015}.
If a Golay pair of length~$v$ exists, then $v$ is called a Golay number and
similarly if a periodic Golay pair of length~$v$ exists, then $v$ is called a periodic Golay number.
A periodic Golay number must be an even number that when doubled is a sum of two squares.
This implies that 14 is not a periodic Golay number, since 28 is not a sum of two squares.

A Golay pair is also a periodic Golay pair, 
but the converse is not necessarily true, as there are periodic Golay numbers that are not Golay numbers.
Currently, the only seven known periodic Golay numbers that are definitely not Golay numbers are 34, 50, 58, 68, 72, 74, and~82~\cite{NA:2015}.
In 1998, Ðoković introduced a canonical form for the equivalence of Golay pairs~\cite{MR1637705}.
A description of all Golay pairs for lengths $v \le 100$ was reported by Borwein and Ferguson in 2003~\cite{Borwein2003ACD}, where five primitive Golay pairs were identified from which all Golay pairs of lengths $v \le 100$ can be derived.
In 1974, Turyn~\cite{Turyn1974} described a method of constructing Golay pairs of length $vm$ from existing pairs of lengths~$v$ and~$m$.
In 2008, Fiedler, Jedwab, and Parker~\cite{MR2450821} used Turyn's construction to demonstrate a framework for Golay pairs of length $2^v$ and also addressed the extra Golay pairs found in exhaustive searches previously conducted by Li and Chu in 2005~\cite{MR2298554}. Fiedler, Jedwab, and Parker also created their own approach of generating Golay pairs in 2006~\cite{MR2417020}, where they consider Golay complementary sequences as projections of multi-dimensional Golay arrays.
Turyn's construction was also slightly generalized to periodic Golay pairs by Đoković and Kotsireas in 2015~\cite{DK:2015:proc} by noticing that Turyn's multiplication formula for Golay pairs can also be used to construct periodic Golay pairs as well. 

Periodic Golay pairs are well-studied, with a number of previous works generating
periodic Golay pairs computationally. 
For example, in 2008, the existence of periodic Golay pairs in lengths $v \le 50$ was settled by Kotsireas et al.~\cite{Kotsireas2008}.
In 2015, eight inequivalent pairs of length~$72$ were generated by Ðoković and Kotsireas~\cite{DK:2015:proc},
and they also generated a pair of length~$74$ in 2014~\cite{NA:2015}.
Our methods not only recreate most of these results but also exhaustively enumerate the lengths, providing complete lists of periodic Golay pairs for lengths $v \le 72$.
Our results confirm and extend the previous known exhaustive enumerations of periodic Golay pairs of lengths $v \le 32$ by Balonin and Đoković in 2015~\cite{Balonin2015} and lengths $v \le 34$ by Crnković, Danilović, Egan, and Švob in 2022~\cite{Dean:2022}.
Previously, the smallest undecided periodic Golay number was $90$~\cite{Dean:2022},
but we provide the first ever examples of periodic Golay pairs of length~$90$ in Section~\ref{sec:PG90}.
Now, the smallest undecided periodic Golay number is $106$~\cite{DK:2015:proc}.

Our search pipeline performs three sequential steps to exhaustively enumerate periodic Golay pairs $(A,B)$ of length $v$: candidate generation, candidate matching, and equivalence filtering.
First,
we generate a list of candidate sequences corresponding to all possible $A$ sequences, and a list of all possible~$B$ sequences (see Sections~\ref{sec:intpart} to~\ref{sec:hybrid}).
Second, we describe an algorithm to efficiently match and test all combinations of $(A, B)$ to construct our final list of periodic Golay pairs (see Section~\ref{sec:matching}).
Third, the last step is to filter our final list up to equivalence, removing
any pairs that are equivalent under the set of
periodic Golay pair equivalence operations (see Section~\ref{sec:equiv}).
This final step reduces the size of the list and provides an experimental verification of results
between independent exhaustive searches.
We provide several algorithms for equivalence filtering in Section~\ref{sec:equivfilter}.

Lengths $v \le 50$ are possible to enumerate using just the generate-and-match method,
but larger orders quickly become unfeasible.
To work around this, we apply sequence compression, introduced in the context
of complementary sequences by Ðoković and Kotsireas~\cite{DK2015},
as a method of reducing the length of candidate sequences at the cost of increasing the alphabet size (see Section~\ref{sec:compression}).
This results in a vastly reduced search space and an improvement in computation time by orders of magnitude.
The downside is that it requires an efficient uncompression method---we cover our uncompression algorithm in Section~\ref{sec:uncompression}.
The notable previous work of Crnkovi\'{c} et al.~\cite{Dean:2022} for enumerating lengths $v \le 34$ also utilized sequence compression, however they noted that the enumeration for length~40 as out of reach with current methods.
In order to go farther, we introduce a method of ``multi-level'' sequence compression
that dramatically improves cases where the length~$v$ is a product of many small primes such as 2 and 3.
Additionally, we make use of the methodology of orderly generation, an algorithm used to generate objects in an isomorph-free way.
This significantly improves the candidate generation step of our pipeline, and can also benefit cases which cannot utilize sequence compression.
We provide precise descriptions of our algorithms throughout the paper,
as well as benchmarks of our implementations on various lengths in Section~\ref{results}.

As a consequence of our extended enumerations, we discover patterns exhibited by our exhaustive data for periodic Golay pairs of larger lengths.
We propose a conjecture for constructing periodic Golay pairs for the lengths $v$ in which periodic Golay pairs exist (see Section~\ref{sec:conjecture}).
We prove that the conjecture holds in all lengths $v < 100$ with the possible exception of $v=50$, and we
experimentally verify that the conjecture does hold for $v=50$.
The conjecture implies a formula exists describing all possible $(v/2)$-compressions of periodic Golay pairs.
We also propose heuristics for the form of $d$-compressions of periodic Golay pairs for compression factors $d$ other than $v/2$
and we demonstrate the usefulness of our heuristics by using them to compute the first ever examples of periodic Golay pairs of length 90.

To our knowledge, our work is the first to search
for complementary sequences using multi-level compression,
but this technique can be considered a special case of
subgroup contraction from the study of difference sets.
To describe a difference set, suppose $D$ is a subset of size~$k$ of a group~$G$ of order $v$.
Then $D$ is a $(v,k,\lambda)$ \emph{difference set}
when the expressions $d_1d_2^{-1}$ for $d_1$, $d_2\in D$ represent every nonidentity element in $G$ exactly $\lambda$ times.
Associating $D$ with $\sum_{d\in D}d$ in the group ring $\mathbb{Z}[G]$,
this can conveniently be expressed by the equation
$DD^{(-1)}=\lambda (G-1)+k$ where $D^{(-1)}=\sum_{d\in D}d^{-1}$.
Periodic Golay pairs are equivalent to
\emph{supplementary difference sets}
in the cyclic group $C_v$, consisting of two subsets $A$ and $B$ of $C_v$
with $AA^{(-1)}+BB^{(-1)}=\lambda C_v+v/2$, where $\lambda=\lvert{A}\rvert+\lvert{B}\rvert-v/2$.
One can convert a periodic Golay pair $([a_0,\dotsc,a_{v-1}],[b_0,\dotsc,b_{v-1}])$ into a supplementary difference set
by taking $A=\{\,g^i:\text{$i$ where $a_i=1$}\,\}$ and $B=\{\,g^i:\text{$i$ where $b_i=1$}\,\}$, where $g$ is a generator of $C_v$.

Subgroup contraction has been used in a number of arguments that put restrictions on
(or show the existence or nonexistence of) difference sets for various parameters.
For example, it was used to search for a $(495,39,3)$ difference set~\cite{smith1990},
to put restrictions on $(320,88,24)$ abelian difference sets~\cite{Ma1995},
$(160,54,18)$ nonabelian difference sets~\cite{Alexander2000},
$(96, 20, 4)$ nonabelian difference sets~\cite{abughneim2004nonabelian},
and to rule out the last 7 open cases from Lander's table of possible abelian difference sets with $k \leq 50$~\cite{Iiams1999}.
Contraction has also been applied to multidimensional arrays whose out-of-phase periodic autocorrelations are zero~\cite{Jedwab1992}.
A subgroup contraction analysis can also be used to refine the orbit structures of a block design
with a given automorphism group, and this was used to show that periodic Golay pairs of length 90 do not exist
under the assumption of some conditions on their automorphism group~\cite{Dean:2022}.

\section{Background on Periodic Golay Pairs}\label{sec:background}

Generally, \emph{complementary sequences} are defined as a collection of sequences such that their periodic or aperiodic autocorrelation functions
sum to a constant value at all shifts $s \neq 0$.
A periodic Golay pair of even length $v$, denoted by $\PG(v)$,
is a pair of two complementary sequences of length $v$, with $\{\pm1\}$ entries,
such that their periodic autocorrelation function coefficients sum to $0$.
Precisely, a periodic Golay pair $(A, B)$ of length $v$ is characterized by the two sequences
\begin{align*}
A &= [a_0,\ldots,a_{v-1}], \text{ where } a_i \in \{\pm1\} \text { for } i=0,\dotsc,v-1, \text{ and} \\
B &= [b_0,\ldots,b_{v-1}], \text{ where } b_i \in \{\pm1\} \text{ for } i=0,\dotsc,v-1,
\end{align*}
as well as the equation
\[
\PAF(A,s) + \PAF(B,s) = 0, \text{ for } s=1,\dotsc,v/2 ,
\]
where $s$ denotes the shift variable, and $\PAF$ denotes the \emph{periodic autocorrelation function} defined by
\[
    \PAF(A,s) = \displaystyle\sum_{k=0}^{v-1} a_k a_{k+s} .
\]
In this definition, the entries of $A$ are taken to repeat periodically, i.e., $a_{v+i}=a_i$ for all $i\geq0$.
A consequence of the definition is that the following sum-of-squares Diophantine equation must have solutions
for a periodic Golay pair of even length $v$ to exist~\cite{NA:2015}.
Precisely, we have the following theorem.

\begin{theorem}\label{thm:diophantine}%
Suppose $(A, B)$ is a\/ $\PG(v)$.
If\/ $a = a_0 + \cdots + a_{v-1}$ and\/ $b = b_0 + \cdots + b_{v-1}$, then
\[
a^2 + b^2 = 2 v . 
\]
\end{theorem}

As a consequence, many lengths for periodic Golay pairs can be immediately ruled out as impossible.
For example, there are only $24$ lengths $v\leq 100$, namely 2, 4, 8, 10, 16, 18, 20, 26, 32, 34, 36, 40, 50, 52, 58, 64, 68, 72, 74, 80, 82, 90, 98, and 100, whose double can be written as a sum of two squares and therefore
are the only integers which can be a periodic Golay number.

Because periodic Golay pairs have entries that have two possibilities, the total size of the search space for a given length~$v$ is at most $2^v \cdot2^v=4^v$. This complexity can be refined by considering only those pairs satisfying the sum-of-squares equation.
A revised search space complexity is calculated as 
\[
\binom{v}{(v+a) / 2} \times \binom{v}{(v+b) / 2},
\]
where $a$ and $b$ are solutions to the Diophantine equation in Theorem~\ref{thm:diophantine},
since a length-$v$ vector with rowsum~$a$ and $\{\pm1\}$ entries contains $(v+a)/2$ ones.
If there are multiple ways of writing $2v$ as a sum of two squares then the size
of the search space needs to account for each decomposition, i.e., the entire
space will be $\sum_{a,b\geq0:a^2+b^2=2v}\binom{v}{(v+a) / 2}\binom{v}{(v+b) / 2}$.

Except for small values of~$v$, using brute force to enumerate this search space is numerically infeasible.
One can reduce the search space from $\binom{v}{(v+a)/2}\binom{v}{(v+b)/2}$ to $\binom{v}{(v+a)/2}+\binom{v}{(v+b)/2}$
at the cost of caching the vectors of $\PAF$ values from all $\{\pm1\}$-vectors with rowsum~$a$, and then for each vector of rowsum~$b$
checking if its $\PAF$ values are the negation of any of the cached $\PAF$ vectors.
However, this would require excessive amounts of memory for the values of~$v$ that we will search.
Thus, we introduce some strong filtering methods to reduce the size of the search space.
One such filtering method is based on a well-known connection between the \textit{discrete Fourier transform} (DFT) of a sequence and its $\PAF$ values.
The $\DFT$ of a sequence $A = [a_0, \dotsc, a_{v-1}]$ at shift $s$ is defined as
\[
    \DFT(A,s) \coloneqq \displaystyle\sum_{k=0}^{v-1} a_k \omega^{ks} ,
\]
where $\omega = e^{2\pi \i / v}$ is a primitive $v$th root of unity.
The \emph{power spectral density} ($\PSD$) of a sequence is defined as the squared magnitude of its $\DFT$ values, i.e.,
\[
    \PSD(A,s) \coloneqq \mathopen|\DFT(A,s)\mathclose|^2 .
\]
When $\PSD(A)$ is written without a shift variable, we mean
the vector consisting of all $\PSD$ values, i.e., $[\PSD(A,0),\dotsc,\PSD(A,v-1)]$ (and similarly
for $\PAF(A)$).

Fletcher, Gysin, and Seberry~\cite[Thm.~2]{Seberry} show that the $\PAF$ values of two sequences sum to a constant if and only if their $\PSD$-values sum to a constant.
Ðoković and Kotsireas~\cite[Thm.~2]{DK2015} prove the
following theorem providing a formula relating the $\PSD$ and $\PAF$ constants of complementary sequences.
\begin{theorem}
\label{thm:paftopsd}
Let $A_1$, $\dotsc$, $A_t$ be complex sequences of length $v$.
These sequences are complementary, i.e., 
\[
\sum_{i=1}^{t} \PAF(A_i) = [\alpha_0, \alpha, \dotsc, \alpha]
\]
if and only if
\[
\sum_{i=1}^{t} \PSD(A_i) = [\beta_0, \beta, \dotsc, \beta] ,
\]
where the $\PAF$ constants $\alpha_0$ and $\alpha$ and the $\PSD$-constants $\beta_0$ and $\beta$ are related by
\[
\beta_0 = \alpha_0 + (v-1)\alpha, \quad\text{and}\quad \beta = \alpha_0 - \alpha .
\]
\end{theorem}
For the case of periodic Golay pairs specifically, where the $\PAF$ constants sum to $0$, this reduces to the following.
\begin{corollary}
\label{cor:paftopsd}
For any two $\{\pm1\}$-sequences $(A, B)$ of even length $v$,
\[
\PAF(A,s) + \PAF(B,s) = 0, \text{ for } s=1,\dotsc,v/2 ,
\]
if and only if
\[
\PSD(A,s) + \PSD(B,s) = 2v, \text{ for } s=0,\dotsc,v/2.
\]
\end{corollary}

This enables us to leverage the $O(v\log v)$ complexity of the fast Fourier transform
rather than the $O(v^2)$ complexity of computing the vector of $\PAF$ values naively.
Furthermore, the $\PSD$ values of a sequence are nonnegative by definition.
This allows us to use the following filtering bound on any individual binary sequence $A$.
\begin{corollary}
Suppose $A$ is a member of a periodic Golay pair of length $v$. Then
\[
    \PSD(A,s) \le 2v.
\]
\end{corollary}
In practice, this filters a large portion of our search space of $2^v$ sequences
and increases the length~$v$ for which it is tractable to enumerate all sequences
that could be a member of a periodic Golay pair.
In certain cases, additional restrictions on permissible
$\PSD$ values of complementary sequences can be derived.
For example, Kotsireas and Koutschan~\cite{MR4373597} prove
that if $(A,B)$ is a Legendre pair (a complementary sequence pair with a $\PAF$ constant of $\alpha=-2$)
then $\PSD(A,v/3)$ and $\PSD(B,v/3)$ are perfect integer squares when $v/3$ is an integer.
The following lemma is an analogue of their result for periodic Golay pairs.
\begin{lemma}
If\/ $A$ is a sequence in a $\PG(v)$,
then $\PSD(A, v/2)$ is a perfect integer square,
and\/ $\PSD(A,v/4)$ is an integer when $v/4$ is an integer.
\end{lemma}
\begin{proofalt}
Consider a primitive $v$th root of unity,      
$\omega=\exp(2\pi \i/v)$, and note that $\omega^{v/2} = -1$.
Thus,
$\PSD(A,v/2)=\bigl|\sum_{k=0}^{v-1}(-1)^k a_k\bigr|^2=\bigl(\sum_{k=0}^{v/2-1}(a_{2k}-a_{2k+1})\bigr)^2$,
the square of an integer.

Note that $\omega^{v/4}=\i$ and
$\PSD(A,v/4) = \DFT(A,v/4)\cdot\overline{\DFT(A,v/4)}$ where the overline denotes
complex conjugation.  Since
\[ \DFT(A,v/4) = \sum_{k=0}^{v/4-1}\bigl(a_{4k}-a_{4k+2}+\i(a_{4k+1}-a_{4k+3})\bigr) = A_0 +\i A_1 \]
where $A_0\coloneqq\sum_{k=0}^{v/4-1}(a_{4k}-a_{4k+2})$ and $A_1\coloneqq\sum_{k=0}^{v/4-1}(a_{4k+1}-a_{4k+3})$ are integers,
we have $\PSD(A,v/4)=(A_0+\i A_1)(A_0-\i A_1)=A_0^2+A_1^2$ is an integer.
\qed
\end{proofalt}
Since $\PSD(A, s) + \PSD(B, s) = 2v$ for all $s$ including $s=v/2$, 
the values $\PSD(A,v/2)$ and $\PSD(B,v/2)$
give a sum-of-squares decomposition of~$2v$---and
all ways of writing~$2v$ as a sum of two integer
squares can be determined in advance.
For example, for $v=68$, up to order and signs the only sum-of-squares decomposition of~$2v$
is $6^2 + 10^2$.  Thus, if $(A,B)$ is a $\PG(68)$, then
$\PSD(A, 34) = 6^2$ and $\PSD(B, 34) = 10^2$ or vice versa.

Despite the large number of sequences that can be filtered by these methods,
the size of the search space for lengths $v > 40$ becomes difficult to cope with.
Thus, a key component of improving the search for a $\PG(v)$ is to reduce the search space
even farther.  We now introduce two tools we used in our searches.

\subsection{Equivalence}\label{sec:equiv}
There are several equivalence operations that can be defined on periodic Golay pairs.
If a pair of sequences $(A, B)$ is a $\PG(v)$, then the following operations can be applied to generate another $\PG(v)$ which is said to be \emph{equivalent} to the original pair.
In the following, $v$ is the length of each sequence in the original sequence pair $(A,B)$,
and $i$ is bounded as $0 \leq i < v$.

\begin{enumerate}
\item{} [Swap] $(B, A)$, swap sequences $A$ and $B$.
\item{} [Shift A] $([a_{i+1 \bmod v}], B)$, rotate the elements of sequence $A$ by one shift.
\item{} [Reverse A] $([a_{(v-1)-i}], B)$, reverse the elements of sequence $A$.
\item{} [Decimation] $([a_{ki \bmod v}], [b_{ki \bmod v}])$, for all $k < v$ with $k$ coprime to $v$, replace $a_i$ with $a_{ki \bmod v}$ and $b_i$ with $b_{ki \bmod v}$.
\item{} [Alternating Negation] $([(-1)^ia_{i}], [(-1)^ib_{i}])$, negate every odd indexed element of sequence $A$ and $B$.
\end{enumerate}

Note that negation, the operation of replacing $(A,B)$ with $(-A,B)$, could also be considered
an equivalence operation.  However, it is not strictly needed as it can be recovered from the above
equivalence operations by applying alternating negation twice with a shift in between.

\subsection{Compression}\label{sec:compression}

Considering the exponential $2^v$ size of the search space
for sequences in a $\PG(v)$, it is useful to reduce the exponent.
Fortunately, one can compress a sequence to a smaller length while maintaining its complementarity~\cite{DK2015}.
Specifically, we can create a new sequence with length $d = v/m$, where $v$ is the length of the original sequence $A = [a_0, \dotsc, a_{v-1}]$, and $m$ is any divisor of $v$.
We call this new sequence $A^{(m)} = [a^{(m)}_0, \dotsc, a^{(m)}_{d-1}]$ the $m$-compression of $A$, with entries that are defined by $a^{(m)}_i = a_i + a_{i + d} + \dotsb + a_{i + (m-1)d}$.
This compression method reduces the size of the search space to $(m+1)^d$. For example, if we compress a length $10$ sequence over a $\{\pm1\}$ alphabet (i.e., a search space of size $2^{10} = 1024$)
by a factor of 2, the compressed search is over
sequences of length $5$ and a $\{\pm2,0\}$ alphabet (i.e., a search space of size $3^5 = 243$).
We use the fact that compressions of a $\PG(v)$ will also be complementary sequences (over a larger alphabet)
in order to search for smaller-length sequences in a reduced search space.
After filtering as many smaller-length sequences as possible,
we apply an uncompression
method to recover periodic Golay pairs of length~$v$.

\section{Candidate Generation}\label{sec:candgen}

Candidate generation is a necessary step in constructing our compressed pairs and in practice is the hardest step of our pipeline due to the aforementioned exponential search space.
Thus, in creating algorithms for generating individual candidate sequences, we strive to consistently reduce the search space by as large a factor as possible to mediate this problem.

\subsection{Integer Partitions}\label{sec:intpart}

There are many characteristics that periodic Golay pairs exhibit that can be used to limit the search space.
For example, as described in Theorem~\ref{thm:diophantine}, the sum of all elements in a given sequence must be a particular value for the sequence to be part of a periodic Golay pair. 
Since it is simple to enumerate all decompositions of $2v$ as a sum of squares, with this information we can limit the search space to only sequences whose rowsums correspond to either $a$ or $b$ where $a^2 + b^2 = 2v$.

To generate only those sequences which sum to $a$ or $b$, the process is straightforward for the case where the given alphabet is $\{-1, +1\}$.
An integer can be partitioned into a sequence of $v$ elements from an alphabet of $\{-1, +1\}$ in exactly one way (up to ordering).
Thus, we can directly count the number of $-1$s and $+1$s that must occur;
there are exactly $(a+v)/2$ positives and $(a-v)/2$ negatives in a $\{\pm1\}$-sequence of length $v$ with rowsum $a$.
However, the problem is more complicated when considering an arbitrary alphabet, as there may be more than one combination of the alphabet that sum to $a$.
After calculating all partitions of $a$ or $b$ into sums of terms from the necessary alphabet (e.g., for 2-compression the alphabet is $\{0,\pm2\}$),
we only consider sequences that are permutations of our partitions, and this gives us our search space.

\subsection{Orderly Generation}\label{sec:ordgen}

Introduced independently by Faradžev \cite{faradvzev1978constructive} and Read \cite{Read1978} in 1978, ``orderly'' generation is a form of isomorph rejection that filters equivalent nodes in a search tree.
Another method of isomorph rejection is the ``canonical augmentation'' method of McKay~\cite{McKay1998}.
By considering just equivalent shifts of a sequence, orderly generation on sequences of length~$v$ can reduce the search space size by approximately a factor of~$v$.

We can theoretically reduce our search space to only inequivalent sequences by limiting our search to only a canonical representative of each equivalence class in our search space.
For periodic Golay pairs, we take the lexicographically minimal sequence of each equivalence class as our canonical representative.

\begin{example}
 Consider the sequence $A = [1, *, *, *]$, where $*\in\{-1, 1\}$ denotes undefined entries.
 If any of the undefined entries are $-1$, then $A$ cannot be a lexicographically minimal representative of its equivalence class,
 since otherwise shifting its entries would produce an equivalent sequence with a first entry of\/ $-1$.
\end{example}
 
This implies that we can always reduce our search space in length~$v$ from $2^v$ to $2^{v-1} + 1$, essentially half of the original size.
But this only removes one branch of the search tree.
To further remove entire sections of our search space during generation, we recursively filter out partially defined subsequences (rather than waiting to filter full sequences of length $v$).
To do this, we generate our sequences using backtracking, starting from the $0$th element, and working down to the $(v-1)$th element.
If it can be shown that any subsequence cannot be lex minimal no matter the value of its remaining elements, then we throw away that branch of the tree and skip to the next.
In practice, this method reduces our search tree to its canonical representatives.
Still, it introduces some computational overhead in calculating whether a subsequence can be lexicographically minimal, typically by generating its equivalence class and determining if a lexicographically smaller subsequence exists.
Furthermore, the number of sequences in a removed branch may be inconsequential.
In practice, we used orderly generation on partially defined sequences with up to $v/2$ defined entries,
as past that point the number of sequences removed is inconsequential.

\subsection{Hybrid Generation}\label{sec:hybrid}

\begin{table}
\begin{tabular}{ccccc}
\toprule
 Length & Compression Factor & Integer Partition & Orderly & Hybrid \\
\midrule
26 & N/A & 4s & 1s & 0s \\
32 & N/A & 264s & 76s & 34s \\
34 & N/A & 1064s & 285s & 120s \\
40 & 2 & 118s & 85s & 38s \\
\botrule
\end{tabular}
\caption{Running time comparison of candidate generation methods. Times are given in seconds.}
\label{tbl:benchcand}
\end{table}

As shown in Table~\ref{tbl:benchcand}, the orderly method outpaces the integer partition method described in Section~\ref{sec:intpart}.
However, there is inefficiency when considering that the further the subsequences are completed, the smaller the removed branches become.
To optimize the stretches of computation where the lower end of a sequence is being completed, we devised a hybrid approach.
Our incomplete sequences are filtered to the length $v/2$, but the remaining length is 
instead constructed using the integer partitions method to consider only the sequences that can sum to $a$ or $b$.
Specifically, our sequences use orderly generation for subsequences up to the length $v/2$.
To complete our subsequences, we want to ignore those sequences that cannot sum to $a$ or $b$, so we compute the current sum of elements of each of our subsequences,
and only fill the rest of our sequence with integer partitions that can complete the sum to $a$ or $b$.
This method takes advantage of both the orderly and integer partitions methods and combines them to complement each other.
As demonstrated in Table~\ref{tbl:benchcand}, this ``hybrid'' method outpaces
both individual methods.

\begin{algorithm}
\caption{$\Match(v,\Omega_A,\Omega_B)$, a sequence matching algorithm.}
\begin{algorithmic}
\State \textbf{Input:} $v$, the length of the sequences to match.\\
$\Omega_A$, a list of sequences which satisfy the $\PSD$ filtering test with constant rowsums. \\
$\Omega_B$, a list of sequences which satisfy the $\PSD$ filtering test with constant rowsums.

\State \textbf{Output:} 
$\Omega$, the list of compatible periodic Golay pairs found from $\Omega_A$ and $\Omega_B$. \\

\item[1.][Sorting]
\State \com Below $\PSD(A)$ and $2v-\PSD(B)$ are defined as the sequences constructed from the values $\PSD(A, s)$ and $2v - \PSD(B, s)$, respectively, for $s = 0, \dots, v/2$.
\State Sort list $\Omega_A$ lexicographically by the values of $\PSD(A)$

\State Sort list $\Omega_B$ lexicographically by the values of $2v-\PSD(B)$ \\

\item[2.][Initialization]

\State $A \coloneqq$ the first sequence in $\Omega_A$

\State $B \coloneqq$ the first sequence in $\Omega_B$

\State $\Omega\coloneqq\emptyset$ \\

\item[3.][Matching]

\While {$A$ is not the end of $\Omega_A$ and $B$ is not the end of $\Omega_B$}

\If{$\PSD(A) = 2v-\PSD(B)$} \quad\com Match found

\State \com Find all sequences with the same PSDs as $A$ and $B$
\State $\text{temp}_A \coloneqq \PSD(A)$
\State $\text{temp}_B \coloneqq 2v-\PSD(B)$
\State $\Gamma_A \coloneqq \emptyset$
\State $\Gamma_B \coloneqq \emptyset$

\While{$A$ is not the end of $\Omega_A$}

\If{$\PSD(A) = \text{temp}_A$}
\State $\Gamma_A \coloneqq \Gamma_A \cup A$
\State $A \coloneqq$ the next sequence in $\Omega_A$
\EndIf

\EndWhile

\While{$B$ is not the end of $\Omega_B$}

\If{$\PSD(B) = \text{temp}_B$}
\State $\Gamma_B \coloneqq \Gamma_B \cup A$
\State $B \coloneqq$ the next sequence in $\Omega_B$
\EndIf
    
\EndWhile

\State $\Omega \coloneqq \Omega \cup (\Gamma_A\times\Gamma_B)$ \quad\com Add all pairs of combinations from $\Gamma_A$ and $\Gamma_B$

\EndIf
\If{$\PSD(A) < 2v - \PSD(B)$} \quad\com Lexicographical comparison
\State $A \coloneqq$ the next sequence in $\Omega_A$
\Else
\quad\com Thus $\PSD(A) \geq 2v-\PSD(B)$
\State $B \coloneqq$ the next sequence in $\Omega_B$
\EndIf

\EndWhile

\end{algorithmic}
\label{algmatching}
\end{algorithm}

\subsection{Matching}\label{sec:matching}

By this point in our pipeline, we are left with a list of $A$ sequences and a list of $B$ sequences
which all pass the $\PSD$ filtering and Diophantine equation tests.
To complete our pipeline, we will use these lists to ``match'' all pairs that satisfy the PSD equation in Corollary~\ref{cor:paftopsd}.
This matching step gives us an exhaustive list of periodic Golay pairs (or, if compression is used, a list of
all compressions of periodic Golay pairs).
The naive solution would be to try all possible combinations of list $A$ and list $B$, but this results in a time complexity quadratic in the list lengths.
We instead use the matching method originally described by Kotsireas, Koukouvinos, and Seberry~\cite{kotsireas2009}, and later used by
Bright, Kotsireas, and Ganesh to exhaustively generate Williamson sequences~\cite{Bright2020}.
This method effectively reduces our matching problem to a string sorting problem.
From Corollary~\ref{cor:paftopsd} we have
\[
\PSD(A,s) = 2v - \PSD(B,s), \text{ for } s = 0, \dots, v/2.
\]
We sort the $A$ sequences so that the values of $\PSD(A,s)$ occur in lexicographic order
and sort the $B$ sequences so the values of $2v - \PSD(B,s)$ occur in lexicographic order.
Then a ``match'', a sequence pair $(A,B)$ with $\PSD(A,s) = 2v - \PSD(B,s)$, corresponds to a periodic Golay pair.
In practice, we round the $\PSD$ values to the nearest integer.
This could introduce spurious solutions, but
when a match occurs it can easily be checked that it is indeed a periodic Golay pair.
The matching process requires only one scan through both lists, making it a linear procedure once the lists have been sorted.
The worst case of this algorithm occurs when every sequence in list $A$ is compatible with every sequence in list $B$.
This case is necessarily quadratic, since the output size is quadratic.
However, this case does not occur in practice for our exhaustive searches.
The precise implementation of this algorithm is given in Algorithm~\ref{algmatching}.

\section{Uncompression}\label{sec:uncompression}

As described in Section~\ref{sec:compression}, the compression method efficiently reduces our initial search space but requires an extra uncompression step.
Uncompression requires constructing all possible sequences of length~$v$ that compress to a given sequence of length $d = v/m$.
This can be an extremely arduous task given a larger compression factor of $m$, so the choice of divisor is very important for efficient generation.
We propose a method of optimizing some cases of large-divisor compression to significantly improve cases where the length of a sequence has many small divisors.

 \subsection{Multi-level Compression}\label{subsec5}

Generating compressed sequences can be difficult since it must be accomplished using a candidate generation step, which can be slow.
Thus, we implemented a new method where we generate candidate $d$-compressed pairs by uncompressing candidate $(d\cdot e)$-compressed pairs
for some uncompression factor $e>1$.
Essentially, we utilize the observation that uncompression is significantly faster with smaller divisors and the fact that compression is also possible with several small divisors, rather than one large divisor.

\begin{theorem}
Suppose a sequence $A$ has length $v$.
For all positive integers $d$, $e$ where $de$ divides $v$,
the $e$-compression of the $d$-compression of $A$, $(A^{(d)})^{(e)}$,
equals $A^{(de)}$, the $(d\cdot e)$-compression of $A$.
\end{theorem}
\begin{proofalt}
Let $A^{(d)}$ denote the $d$-compression of $A$ and with entries given by
$a_k^{(d)}=\sum_{i=0}^{d-1} a_{k+iv/d}$ for $k=0$, $\dotsc$, $v/d-1$.
The $k$th entry of the $e$-compression of the $d$-compression of $A$ is
\[
(A^{(d)})^{(e)}_k = \sum_{i=0}^{e-1} a^{(d)}_{k+i(v/d)/e}
= \sum_{i=0}^{e-1} \sum_{j=0}^{d-1} a_{k+(i+ej)v/(de)}
= \sum_{i=0}^{de-1} a_{k+iv/(de)} = a^{(de)}_k
\]
where the third equality rewrites the double sum as a single sum using the identity
$\bigcup_{i=0}^{e-1}\bigcup_{j=0}^{d-1}\{i+ej\}=\{0,\dotsc,de-1\}$. \qed
\end{proofalt}

This allows us to perform uncompression by a large compression factor by splitting it into several rounds of uncompression by intermediate factors.

\subsection{Description of Uncompression}\label{sec:descuncomp}

An uncompression algorithm must generate a list of uncompressions with the requirement that
if a sequence compresses to a given candidate compression,
then that sequence must be present in the list of uncompressions.
In other words, we need to compute the preimage of a given sequence under compression.
A single entry in a compression involves a fixed-length sum,
and this implies that to revert this process and uncompress a sequence,
we need to calculate all possible fixed-length sums for each element
and construct all possible combinations of these for our final list of sequences.

\begin{example}
If we are to uncompress the sequence $A = [0, 2]$, with compression factor~$2$,
the uncompressed list must be $[-1, 1, 1, 1]$ or $[1, 1, -1, 1]$, by observing that $0 = -1 + 1$ and $2 = 1 + 1$
are the only possible length-$2$ $\{\pm1\}$-sums of $0$ and $2$.
\end{example}

Since all possible permutations of these fixed-length partitions must be considered, we utilize backtracking to implement the algorithm. Utilizing recursion to implement backtracking, the detailed algorithm is described in Algorithms~\ref{uncompressioncall} and~\ref{uncompressionalg}.
\begin{algorithm}
\caption{$\Uncompress(\omega, e, d)$, uncompress a sequence $\omega$ by a factor of $e$ (a wrapper for Algorithm~\ref{uncompressionalg}).}
\begin{algorithmic}
\State \textbf{Input:}
$\omega$, the original compressed sequence with length $v$.\\
$d$, the compression factor of the sequences which $e$-compress to $\omega$.\\
$e$, the uncompression factor to be applied to $\omega$.

\State \textbf{Output:} 
The set of all length $ve$ sequences that $e$-compress to $\omega$. \\

\item[1.][Initialization]

\State ${\gamma} \coloneqq $ A sequence of length $m \coloneqq ve$ with all undefined elements.

\State \Return $\Uncompress(\gamma, \omega, 0, d, e)$ \quad\com All sequences that $e$-compress to $\omega$

\end{algorithmic}
\label{uncompressioncall}
\end{algorithm}
\begin{algorithm}
\caption{$\Uncompress(\gamma, \omega, i, d, e)$, uncompression algorithm with recursive backtracking.}\label{alg:cap}
\begin{algorithmic}
\State \textbf{Input:}
$\omega$, the original compressed sequence of length $v$.\\
$\gamma$, a sequence of length $m \coloneqq ve$ with $ie$ defined entries and $(v-i)e$ undefined entries. \\
$i$, the index of $\omega$ at which to start uncompressing.\\
$d$, the compression factor of the sequences which $e$-compress to $\omega$.\\
$e$, the uncompression factor to be applied to $\omega$.

\State \textbf{Output:} $\Omega$, the set of all length $m$ sequences
whose entries match the \emph{defined} entries of $\gamma$
and whose other entries $e$-compress to $[\omega_i, \dotsc, \omega_{v-1}]$. \\

\item[1.][Initialization]

\State $\Omega \coloneqq$ an empty set of length $m$ sequences.

\State \com We are to construct permutations that account for all possibilities of sequences that could compress to $\omega_i$. These permutations consist of elements in the set $A \coloneqq \{\,x \in \mathbb{Z} :
-d \le x \le d \text{ and } x \equiv d \pmod{2}\,\}$

\State $\Omega_\text{subseq} \coloneqq$ the set of all length-$e$ integer permutations of the $i$th element of $\omega$, $\omega_i$.  

\\
\item[2.][Base case, nothing to uncompress]

\If{$i$ == $v$} 
 \State \Return $\{\gamma\}$
\EndIf

\\
\item[3.][Recursive step, uncompress all possible permutations of $\omega_i$ into $\gamma$]\\
 
\com Loop to construct subsequences using all possible length-$e$ integer permutations of $\omega_i$

\For{$\delta$ in $\Omega_\text{subseq}$}

\State \com Assign the elements of the current permutation into $\gamma$

\For{$j$ in $\{0, \dotsc, e - 1\}$} 
    
\State $\gamma_{i + jv} \coloneqq \delta_j$

\EndFor

\State \com Recursively call the algorithm to construct all combinations

\State $\Omega \coloneqq \Omega \cup \Uncompress(\gamma, \omega, i + 1, d, e)$

\EndFor%

\end{algorithmic}
\label{uncompressionalg}
\end{algorithm}
\section{Equivalence Filtering}\label{sec:equivfilter}

It is convenient to consider only pairs that are mutually inequivalent.
This is useful in optimizing various steps of our algorithms,
as we can allow ourselves to only process those inequivalent pairs, reducing our computational costs by a factor up to $32v^2\varphi(v)$ for length $v$~\cite{Dean:2022}, where $\varphi(v)$ denotes Euler's totient function. 
Furthermore, it is difficult to verify the correctness of the results of different methods of search if the results are not standardized in some form.

The general filtering algorithm described by Bright~et~al.~\cite[Alg.~3.1]{Bright2021} for the postprocessing of complex Golay pairs (Golay pairs over the alphabet $\{\pm1,\pm\i\}$~\cite{Craigen2002}) accomplishes this task.
This algorithm was used for filtering complex Golay pairs, and for periodic Golay pairs it quickly reaches unfeasibility for larger lengths due to its computation and memory requirements.
The method filters sequences by iterating through the list of sequences, computing and caching the equivalence class of every sequence that has not previously been seen in any previous equivalence class.
If a sequence has been found in a previous equivalence class, we remove it.
This method essentially requires storing the set of all periodic Golay pairs in memory, which is unfeasible in our case.
We provide modifications that address these problems. In what follows, $\Omega$ will refer to a set of periodic Golay pairs,
and $\omega$ will refer to a pair of sequences (typically implemented using two arrays).

\subsection{Equivalence Class Generation}

The most straightforward algorithm to generate the equivalence class of a given pair is described in Algorithm~\ref{equivclass}, where we apply our equivalence operations to our given base pair and continuously apply the equivalence operations to all newly generated pairs until no new pairs are found in an iteration.
Our equivalence filtering methods require the generation of equivalence classes of all given pairs, so an efficient method of equivalence class generation is necessary.

\begin{algorithm}[H]
\caption{$\Equiv(\omega_\text{base})$, the first algorithm to generate the equivalence class of a pair $\omega_\text{base}$.}

\begin{algorithmic}

\State \textbf{Input:}
$\omega_\text{base}$, the original sequence pair of length $v$.

\State \textbf{Output:} 
$\Omega_\text{class}$, the equivalence class of $\omega_\text{base}$.

\State \com Note that $\Omega_\text{class}$ changes each iteration
\State $\Omega_\text{class} \coloneqq$ \{$\omega_\text{base}$\}
\For{$\omega \in \Omega_\text{class}$}

\State \com $e_1, \dots, e_n$ is a list of generators of the $\PG(v)$ symmetry group
\State $\Omega_\text{class} \coloneqq \Omega_\text{class} \cup \{e_1(\omega),\dots, e_n(\omega)\} $

\EndFor

\end{algorithmic}
\label{equivclass}
\end{algorithm}

Algorithm~\ref{equivclass} is far too slow for larger lengths $v$, where an equivalence class has a size of up to $32v^2\varphi(v)$~\cite{Dean:2022} because each equivalent sequence is generated several times.
For example, if the equivalence operations $e_1, \dots, e_n$ are performed on a sequence~$\omega$,
it is evident that $\omega$ will be redundantly generated again from the set obtained in $\{e_1(\omega),\dots, e_n(\omega)\}$ directly in the next iteration.
Thus, we develop a faster algorithm to generate equivalence classes.

Let $\omega$ be our original sequence, and let $\Omega_{\text{sym}}$ be the group of symmetries generated from applying Algorithm~\ref{equivclass} on $\omega_{\text{sym}}$, where
\[
    \omega_{\text{sym}} = ([1, 2, \dots, v], [v + 1, v + 2, \dots, 2v]).
\]
Essentially, we construct a sequence $\omega_{\text{sym}}$ with elements consisting of~$2v$ variables, with each variable $\{1, \dots, 2v\}$ representing a distinct index in a
pair of length-$v$ sequences.
Note $\omega_{\text{sym}}$ is not a $\PG(v)$ itself; it is a symbolic representation of the identity equivalence operation
where the entry at each position gets mapped to itself.
In this way, the set $\Omega_{\text{sym}}$ will have size $32v^2\varphi(v)$ and encode each symmetry in the symmetry group of periodic Golay pairs of length~$v$.
Algorithm~\ref{symmetry} describes the procedure for generating an equivalence class from the $\PG(v)$ symmetry group. For each symmetry $\omega_\text{sym}$ in $\Omega_{\text{sym}}$, we create an equivalent sequence $\omega_\text{e}$ from $\omega$.

\begin{algorithm}[H]
\caption{$\Equiv(\omega_\text{base})$, an alternate algorithm to generate the equivalence class of a pair $\omega_\text{base}$.}

\begin{algorithmic}

\State \textbf{Input:}
$\omega_\text{base}$, the original sequence pair of length $v$.

\State \textbf{Output:} 
$\Omega_\text{class}$, the equivalence class of $\omega_\text{base}$.

\State $\Omega_\text{class} \coloneqq$ \{$\omega_\text{base}$\}
\State \com $\Omega_\text{sym}$ is the $\PG(v)$ symmetry group
\For{$\omega_\text{sym} \in \Omega_\text{sym}$}

\State $\omega_\text{e} \coloneqq \textsc{Apply}(\omega_\text{sym}, \omega)$, \text{the symmetry $\omega_\text{sym}$ applied to $\omega$} 

\State $\Omega_\text{class} \coloneqq \Omega_\text{class} \cup \{\omega_\text{e}\} $

\EndFor

\end{algorithmic}
\label{symmetry}
\end{algorithm}

To apply a symmetry $\omega_\text{sym}$ to a sequence $\omega$, we effectively rearrange the indexes of our base pair from the sequences in our symmetry group. Using our particular encoding of the $\PG(v)$ symmetry group, symmetries are applied using the following procedure:

\begin{enumerate}

\item If the $i$th element of $\omega_\text{sym}$ is $k$, then set it to the $(k-1)$th element of $\omega$.

\item If the $i$th element of $\omega_\text{sym}$ is negative, then negate the $(k-1)$th element of $\omega$.

\item If the elements of sequence $A$ of $\omega_\text{sym}$ refer to indexes larger than $v$, then swap the sequences.

\end{enumerate}
We repeat this procedure for each sequence in the symmetry group of periodic Golay pairs, thus directly generating our equivalence class from the symmetry group.
This method avoids the repetition where an equivalent pair is regenerated several times; the method only needs a single iteration over the $32v^2\varphi(v)$ elements
of $\Omega_\text{sym}$.

\subsection{List Filtering}
The output of an equivalence filtering algorithm is a list containing exactly one representative of each equivalence class present in the original list.
The naive method requires storing the equivalence class of every pair in our original list, which is not feasible for lengths $v > 32$.
To filter equivalent pairs from our list of generated pairs, we loop through our list, storing the equivalence class of our current pair, and removing all other pairs from the list that are in this pair's equivalence class.
Our modified algorithm avoids the overbearing memory costs of the original by only storing one equivalence class at a time, rather than caching the entire set of all equivalence classes throughout the computation.
Note that this modified algorithm will have a small but noticeable slowdown due to the added set subtraction step.

\begin{algorithm}[H]
\caption{$\Filter(\Omega_\text{pairs})$, an equivalence filtering algorithm for a set of pairs $\Omega_\text{pairs}$.}\label{alg:listfilt}
\label{equivfiltering}
\begin{algorithmic}

\State \textbf{Input:}
$\Omega_\text{pairs}$, the original list of periodic Golay pairs.

\State \textbf{Output:} 
$\Omega_\text{inequiv}$, a set containing exactly one representative of each equivalence class present in $\Omega_\text{pairs}$.

\State $\Omega_\text{inequiv} \coloneqq \Omega_\text{pairs}$

\For{$\omega \in \Omega_\text{inequiv}$}

\State \com Generate the equivalence class of $\omega$, removing $\omega$ itself
\State $\Omega_\text{class} \coloneqq \Equiv(\omega) \setminus \{\omega\}$

\State \com Remove all sequences equivalent to $\omega$ from $\Omega_\text{inequiv}$, excluding $\omega$
\State $\Omega_\text{inequiv} \coloneqq \Omega_\text{inequiv} \setminus \Omega_\text{class}$

\EndFor

\end{algorithmic}
\end{algorithm}

Algorithm~\ref{equivfiltering} filters a list by iterating through the original list from its beginning to its end.
On each iteration, the current sequence pair $\omega$ is used as a representative of an equivalence class
and the algorithm removes all pairs equivalent to $\omega$ from the list.
It follows that we will be left with exactly one representative of each equivalence class.
This algorithm requires the generation of each equivalence class exactly once,
but it still isn't ideal in all cases due to the amount of memory used by the set operations.
This makes the algorithm difficult to parallelize, and thus unfeasible for larger computations.

\begin{algorithm}[H]
\caption{$\Filter(\Omega_\text{pairs})$, a parallelizable equivalence filtering algorithm for a set of pairs $\Omega_\text{pairs}$.}\label{canon}
\begin{algorithmic}

\State \textbf{Input:}
$\Omega_\text{pairs}$, the original list of periodic Golay pairs.

\State \textbf{Output:} 
$\Omega_\text{inequiv}$, a set containing exactly one representative of each equivalence class present in $\Omega_\text{pairs}$.

\State $\Omega_\text{inequiv} \coloneqq \emptyset$

\For{$\omega \in \Omega_\text{pairs}$}

\State \com Generate the equivalence class of $\omega$
\State $\Omega_\text{class} \coloneqq \Equiv(\omega)$

\State \com Take the lexicographically minimal sequence from $\Omega_\text{class}$ as our representative
\State $\Omega_\text{inequiv} \coloneqq \Omega_\text{inequiv} \cup \{\text{min}(\Omega_\text{class})\}$

\EndFor

\end{algorithmic}
\end{algorithm}

Instead, we modify Algorithm~\ref{alg:listfilt} to generate the equivalence class of all sequences in our original list and take exactly one canonical representative from each to store in our final list.
Algorithm~\ref{canon} converts each pair in the original list to its lexicographically minimal form.
Thus, if two pairs are equivalent, they get the same minimal canonical representative.
We are left with a set containing a canonical representative of each equivalence class.
This algorithm is better for larger lists because it avoids the complexity of subtracting two sets.
The generation of each equivalence class can be done independently, so this algorithm can be trivially parallelized.
The only weakness of this method compared to the first is that if two pairs are equivalent, both of their equivalence classes will be generated.
This problem can be remedied by filtering ``partially'' up to only a few select equivalences, which can be performed very quickly, and then fully filtering this reduced list.
Algorithm~\ref{canon} was necessary instead of Algorithm~\ref{equivfiltering} for computing the cases where our original lists were too large to compute without
parallelization---in particular, the lengths~64 and~68.
With these modifications, and by selecting our equivalences strategically, candidate compression sequences can be postprocessed very quickly in comparison to their generation time.
Furthermore, the exhaustive enumeration of periodic Golay pairs up to all equivalences is made feasible
for all lengths $v\leq72$.

\section{Results}\label{results}

\begin{table}
\begin{tabular}{ccccc}
\toprule
 Order & Candidate Generation & Uncompression & Total Time & Compression Factor \\
\midrule
 20 & 0s & N/A & 0.0d & N/A\\
 26 & 1s & N/A & 0.0d & N/A\\
 32 & 34s & N/A & 0.0d & N/A\\
 34 & 9s & 4s & 0.0d & 2\\
 40 & 38s & 211s & 0.0d & 2\\
 50 & 121621s & 3264s & 1.4d & 2\\  
 52 & 58s & 64810s & 0.8d & 4 $\hookrightarrow$ 2\\
 58 & 1555206s & 180172s & 20.1d & 2\\
 64 & 5s & 126187201s & 1460.5d & 8 $\hookrightarrow$ 4 $\hookrightarrow$ 2\\
 68 & 9292s & 315361452s & 3650.1d & 4 $\hookrightarrow$ 2\\
 72 & 20s & 175812s & 2.0d & 8 $\hookrightarrow$ 4 $\hookrightarrow$ 2\\
\botrule
\end{tabular}
\caption{Timings recorded for the exhaustive searches of lengths $v \le 72$.
Times for candidate generation and uncompression are given in seconds, and the total
time used is given in days.}
\label{timings}
\end{table}

\begin{table}
\begin{tabular}{cccc}
\toprule
 Length & Equivalence Classes & Filtering Time 
 & Filtering Time\\
\midrule
 2 & 1 & 0s & 0.0d\\
 4 & 1 & 0s & 0.0d\\  
 8 & 2 & 0s & 0.0d\\  
 10 & 1 & 0s & 0.0d\\  
 16 & 11 & 1s & 0.0d\\  
 20 & 34 & 7s & 0.0d\\  
 26 & 53 & 35s & 0.0d\\
 32 & 838 & 989s & 0.0d\\ 
 34 & 373 & 655s & 0.0d\\ 
 40 & 9281 & 14732s & 0.2d\\ 
 50 & 8753 & 22427s & 0.3d\\ 
 52 & 14354 & 194653s & 2.2d\\ 
 58 & 13386 & 313832s & 3.6d\\ 
 64 & 1112383  & 518413046s & 6000.1d\\ 
 68 & 90240 & 18041532s & 208.8d\\
 72 & 1753 & 41400s & 0.5d\\
\botrule
\end{tabular}
\caption{Counts for the number of equivalence classes found in each search and filtering time
given in seconds and days.}
\label{equivcount}
\end{table}

All of our algorithms are implemented in C++, and we make use of the FFTW (Fastest Fourier Transform in the West) C library~\cite{Frigo2005} to quickly calculate the $\PSD$ values.
Our results are archived online,\footnote[1]{Archived results available at \url{https://zenodo.org/records/12792345}.}
and to reproduce our results we provide an online repository containing our code as well as instructions for installing and executing our code.\footnote[2]{Code available at \url{https://github.com/tylerlumsden/GolayPair}.}
Our computations were performed on the Compute Canada~\cite{baldwin2012compute} cluster Cedar.
The computations were all performed using no more than 4 GiB of memory and were run on Intel CPUs
of type Broadwell, Cascade Lake, or Skylake.

The timings of our algorithms in practice are provided in Tables~\ref{timings} and~\ref{equivcount}.
Table~\ref{equivcount} also provides the number of equivalence classes found by our searches in each length.
The matching step is omitted due to the inconsequential computation time relative to the rest of the search. Our computations exhaustively searched all lengths $v \le 72$. All lengths 52 and smaller are easily reproducible, as Table~\ref{timings} indicates that they require less than a day of computation time using the multi-level compression method.

The length 58 search could not be done using multi-level compression, as the prime factorization of 58 is $2 \times 29$,
meaning only compression by a factor of 2 or 29 is possible.
For $v=58$, the generation of its candidate 2-compressions took $\approx18$ days, whereas the 2-compression generation in length~64 took only 12 hours, showing the extreme improvement in efficiency achieved by using multi-level compression.
The uncompression step for $v=58$ took about 2 days to compute.

The computations for length 64 were done in parallel, with the final uncompression step taking roughly 4 CPU years.
The length 64 search was completed by first generating all 4-compressions of length $64/4=16$,
then uncompressing those to an intermediate length resulting in 2-compressions of length $32$.
These resulted in roughly $\approx160$ million candidate 2-compressed pairs.
The final step was to uncompress these 2-compressions, which required 4 CPU years, as each uncompression took less than one second.
Every step before the final 2-uncompression took only 12 hours.

The exhaustive search for length 68 required 9292 seconds of candidate generation
of 4-compressions of length $68/4=17$, and 6.25 CPU days to uncompress these to 2-compressed pairs.
Afterward, the final uncompression step took 10 CPU years.

As noted in Table~\ref{equivcount}, the equivalence class count for $v=72$ is significantly smaller than the preceding lengths.
This fact contributed to the short computation time in length 72, as in comparison the length 64 search which had $\approx160$ million candidate 2-compression pairs,
the length 72 search had only 64,601 candidate 2-compression pairs.
The likely reason for the small equivalence class counts is because its direct divisors, 18 and 36, are not periodic Golay numbers.

Our algorithms reached an exhaustive classification for lengths $v \le 72$,
significantly extending the recent classification of periodic Golay pairs for lengths $v \le 34$
by Crnković, Danilović, Egan, and Švob~\cite{Dean:2022}.
For the lengths $v \le 34$, our results are identical to their reported results.
Balonin and Đoković~\cite{Balonin2015} give results
for exhaustive searches that they performed in all lengths $v\leq 26$,
and also report some counts for the number of equivalence classes they found in the
lengths $v\in\{32,34,40\}$.
By personal communication, we were informed that the searches with $v\leq32$ were exhaustive
and the searches with $v>32$ were not exhaustive.
Indeed, our counts match their counts for all lengths $v\leq32$.
For $\PG(34)$, their non-exhaustive search found 256 equivalence classes (117 fewer than our search), but
surprisingly for $\PG(40)$ they reported finding 9301 equivalence classes (20 classes more than our search).
We contacted the authors, and they provided us with representatives of 368 equivalence classes
for $\PG(40)$ and all of these were present in our enumeration.
The source of the 9301 count for $\PG(40)$ is unclear and appears to simply be a misprint.

\section{The $(v/2)$-compression Conjecture for $\PG(v)$}\label{sec:conjecture}

An examination of the exhaustive search results obtained in Section~\ref{results} reveals some structure holding for all periodic Golay pairs
of lengths $v\leq72$.  It also holds for all additional sporadic examples of periodic Golay pairs we
found in the literature~\cite{NA:2015,DK:2015:proc,Dean:2022}.
To describe the structure, suppose
$(A,B)$ is the $(v/2)$-decomposition of a $\PG(v)$ with rowsums $a$ and $b$, i.e., $a^2+b^2=2v$.
All $(v/2)$-compressions of $\PG(v)$ that we considered were equivalent to one of two forms, either
$([0,a],[0,b])$ or $\bigl([\frac{a+b}{2},\frac{a-b}{2}],[\frac{a+b}{2},\frac{b-a}{2}]\bigr)$ and counts for the number of $(v/2)$-compressions in each of the two forms
are presented in Table~\ref{tbl:Conjecture1}.
Note that the case $([0,a],[0,b])$ never occurs
when $v/2$ is odd, a fact we prove in Lemma~\ref{thm:case-elim} below.
We also prove that when $v/2$ is an odd prime that
there is only one possible $(v/2)$-compression up to signs and order---see Theorem~\ref{thm:conjproof}.

\begin{table}
\begin{tabular}{cccc}
\toprule
$v$  & $(a, b)$ & $\bigl([\frac{a+b}{2},\frac{a-b}{2}],[\frac{a+b}{2},\frac{b-a}{2}]\bigr)$ & $([0,a]$, $[0,b])$ \\
\midrule
2 & $(0,2)$ & 1 & 0\\
4 & $(2,2)$ & 1 & 1\\
8 & $(0,4)$ & 1 &  1\\
10 & $(2,4)$ & 1 & 0\\
16 & $(4,4)$ & 11 & 11\\
18 & $(0,6)$ & 0 & 0\\
20 & $(2,6)$ & 21 & 13\\
26 & $(4,6)$ & 53 & 0\\
32 & $(0,8)$ & 400 & 438\\
34 & $(2,8)$ & 373 &  0\\
36 & $(6,6)$ & 0 & 0\\
40 & $(4,8)$ & 6012 & 3269 \\ 
50 & $(6,8)$ & 5293 & 0\\
50 & $(0,10)$ & 3460 & 0\\
52 & $(2,10)$ & 8619 & 5735\\
58 & $(4,10)$ & 13386 & 0 \\
64 & $(8,8)$ & 1112383 & 1112383 \\
68 & $(6,10)$ & 57095 & 33145\\
72 & $(0,12)$ & 863 & 890 \\
\botrule
\end{tabular}
\caption{Counts of $(v/2)$-compressions for all $\PG(v)$ with $v\leq72$.}
\label{tbl:Conjecture1}
\end{table}

\begin{lemma}\label{thm:case-elim}
Suppose $v$ is a periodic Golay number with $v/2$ odd
and let $(a,b)$ be a solution of\/ $a^2+b^2=2v$.
Then $([0,a],[0,b])$ cannot be the $(v/2)$-compression of any $\PG(v)$.
\end{lemma}
\begin{proofalt}
When $v/2$ is odd, the $(v/2)$-compression of any $\{\pm1\}$-sequence only consists of odd entries, as each element is constructed from an odd number of $\pm1$s.
Since $0$, $a$, and~$b$ are all even integers,
they cannot be entries of a $(v/2)$-compression of a $\PG(v)$. \qed
\end{proofalt}

\begin{theorem}\label{thm:conjproof}
Suppose $(A,B)$ is the\/ $(v/2)$-decomposition of a $\PG(v)$ with rowsums $a$ and $b$, i.e., $a^2+b^2=2v$.
If $v/2$ is an odd prime, then $(A,B)$ is equivalent to $\bigl([\frac{a+b}{2},\frac{a-b}{2}],[\frac{a+b}{2},\frac{b-a}{2}]\bigr)$.
\label{Conjecture1}
\end{theorem}
\begin{proofalt}
Denote the entries of $(A,B)$
by $([a_0,a_1],[b_0,b_1])$
and note that $a_0+a_1=a$ and $b_0+b_1=b$.
Because compression preserves complementarity, we have
\[ \PAF(A,1)+\PAF(B,1) = 2a_0a_1 + 2b_0b_1 = 0 . \]
Dividing by $2$ and using $a_1=a-a_0$ and $b_1=b-b_0$, we have
$a_0(a-a_0)+b_0(b-b_0) = 0$ or equivalently
$a_0^2-a a_0 + b_0^2-b b_0 = 0$.
Completing the square twice yields
\[ (a_0-a/2)^2 - a^2/4 + (b_0-b/2)^2 - b^2/4 = 0 , \]
which, as $a^2+b^2=2v$, gives
\[ (a_0-a/2)^2 + (b_0-b/2)^2 = v/2 . \]
Since $a$, $b$, and $v$ are all even, this
is a decomposition of the odd prime $v/2$ into a sum of two integer squares.
It is well known (e.g., see Section~16.9 of \cite{hardy2008introduction}) that
$v/2$ must be of the form $4m+1$ for such a decomposition to exist
and in total there are eight distinct decompositions.
Up to order and signs all decompositions are
equivalent to $(a/2)^2+(b/2)^2=v/2$;
the first four decompositions are given by all choices of signs in
$a_0=a/2\pm a/2$ and $b_0=b/2\pm b/2$,
and the last four are given by all choices of signs in
$a_0=a/2\pm b/2$ and $b_0=b/2\pm a/2$.

The first four cases for $(a_0,b_0)$ result in $(A,B)$ being equivalent to $([0,a],[0,b])$.
Since $v/2$ is odd, by Lemma~\ref{thm:case-elim} these are not possible $(v/2)$-compressions.
The remaining four cases for $(a_0,b_0)$ all result in $(A,B)$ being equivalent to
$\bigl([\frac{a+b}{2},\frac{a-b}{2}],[\frac{a+b}{2},\frac{b-a}{2}]\bigr)$.%
\qed
\end{proofalt}

Note that the proof of Theorem~\ref{thm:conjproof} easily generalizes whenever $v/2$
is either not divisible by a prime of the form $4m+1$,
or its prime factorization has a prime of the form $4m+1$ appearing exactly once.
In that case (and only that case),
there is at most a single decomposition of $v/2$ up to order and signs.

\begin{theorem}\label{thm:extended}
Suppose $(A,B)$ is the $(v/2)$-decomposition of a $\PG(v)$ with rowsums $a$ and $b$.
If the prime factorization of\/ $v/2$ contains no prime of the form $4m+1$, or contains
a single prime of the form $4m+1$ appearing exactly once, then
$(A,B)$ is equivalent to either\/ $([0,a],[0,b])$ or\/
$\bigl([\frac{a+b}{2},\frac{a-b}{2}],[\frac{a+b}{2},\frac{b-a}{2}]\bigr)$.
\end{theorem}

Theorem~\ref{thm:extended} explains the results of every case presented in Table~\ref{tbl:Conjecture1}
with the exception of $v=50$, because the prime factorization of $v/2=5^2$ contains two
copies of the prime~$5$.  In that case, there are two inequivalent
decompositions of $v/2$ into a sum of two integer squares, $(\alpha_1,\beta_1)=(0,5)$,
and $(\alpha_2,\beta_2)=(3,4)$.  Multiple inequivalent decompositions
raises the possibility that there might exist $(v/2)$-compressions of other
forms, such as
$([\alpha_2+\alpha_1,\alpha_2-\alpha_1],[\beta_2+\beta_1,\beta_2-\beta_1])$.
In the case $v=50$ and $(a,b)=(6,8)$, this would result in the $(v/2)$-compression
$([3,3],[9,-1])$, but interestingly
this was not the $25$-compression of any $\PG(50)$ found by our exhaustive search.
The fact that $([3,3],[9,-1])$ didn't occur in practice could potentially indicate
there is a yet-undiscovered theoretical reason why such a $(v/2)$-compression is not possible.
This makes it tempting to conjecture that the condition on the prime factorization
of $v/2$ in Theorem~\ref{thm:extended} might be dropped.

\begin{conjecture}\label{conj:comp}
Suppose $(A,B)$ is the $(v/2)$-decomposition of a $\PG(v)$ with rowsums $a$ and $b$.
Then $(A,B)$ is equivalent to either $([0,a],[0,b])$ or $\bigl([\frac{a+b}{2},\frac{a-b}{2}],[\frac{a+b}{2},\frac{b-a}{2}]\bigr)$.
\end{conjecture}

Given we have experimentally tested Conjecture~\ref{conj:comp} in only a single
length that Theorem~\ref{thm:extended} does not apply to, the evidence for this conjecture
is currently rather weak.  However, even if the conjecture is false it would be interesting to determine
for which other lengths $v$ it holds for.  The first length for which the
correctness of Conjecture~\ref{conj:comp} is uncertain is $v=100$.

We have also found an interesting phenomenon when analyzing the $d$-compressions
of the results of our exhaustive searches for values of $d$ other than $v/2$.
Often, our $d$-compressed pairs exhibit the characteristic that the $\PAF$ values are all zero.
That is, $\PAF(A, s) = 0$ and $\PAF(B, s) = 0$  for $s = 1$, $\dotsc$, $v/2$,
and thus trivially satisfying the $\PAF$ equation of Theorem~\ref{thm:paftopsd}.
We have found examples of this occurring in all lengths~$v$ for which $v/2$ is not prime.
Table~\ref{tbl:Conjecture3} lists the values of $v$ for which we verified this phenomenon.
As shown in the table, there are often many divisors $d$ that work for a specific $v$.
In addition, note that there are (up to sign and order) two possible rowsum possibilities
for $v=50$, and there are $d$-compressions of $\PG(50)$s with all-zero $\PAF$s
using $d=10$ for both rowsum possibilities.

\begin{table}
\begin{tabular}{ccc}
\toprule
$v$ & $d$ & sequences of length $v/d$ with zero PAF that $d$-uncompress to a $\PG(v)$ \\
\midrule
16 & 2 & $[-2, 0, 2, 0, 2, 0, 2, 0]$, $[-2, 0, 2, 0, 2, 0, 2, 0]$  \quad \circledequal \\
20 & 5 & $[1, 1, 1, -1]$, $[3, 3, 3, -3]$ \\
32 & 8 & $[0, 0, 0, 0]$, $[4, 4, 4, -4]$ \\
40 & 4 & $[0, 0, 0, 0, 0, 0, 0, 4, 0, 0]$, $[0, -4, 0, 0, 4, 0, 4, 0, 0, 4]$   \\
40 & 8 &  $[0, 0, 0, 0, 4]$, $[0, 0, 0, 0, 8]$  \\
50 & 10 & $[0, 0, 0, 0, 0]$, $[0, 0, 0, 0, 10]$ \\
50 & 10 & $[0, 0, 0, 0, 6]$, $[0, 0, 0, 0, 8]$  \\
52 & 13 & $[1, 1, 1, -1]$, $[5, -5, 5, 5]$ \\
64 & 4 & $[-4, 0, 0, 0, 0, 0, 4, 0, 4, 0, 0, 0, 0, 0, 4, 0]$, $[-4, 0, 0, 0, 0, 0, 4, 0, 4, 0, 0, 0, 0, 0, 4, 0]$ \quad \circledequal \\
64 & 8 & $[-4, 0, 0, 4, 4, 0, 0, 4]$, $[-4, 0, 0, 4, 4, 0, 0, 4]$ \quad \circledequal \\
64 & 16 & $[0, 0, 0, 8]$, $[0, 0, 0, 8]$ \quad \circledequal \\
68 & 17 & $[5, -5, -5, -5]$, $[3, -3, -3, -3]$ \\
72 & 8 & $[0, 0, 0, 0, 0, 0, 0, 0, 0]$, $[0, -8, 4, 0, 4, 4, 0, 4, 4]$ \\
72 & 12 & $[0, 0, 0, 0, 0, 0]$, $[0, 0, 0, 0, 0, 12]$ \\
72 & 24 & $[0, 0, 0]$, $[0, 0, 12]$ \\
90 & 15 & $[3, 3, 3, 3, -9, 3], [-3, 3, 3, 3, 3, 3]$\\
90 & 18 & $[0, 0, 0, 0, 6], [0, 0, 0, 0, 12]$\\
\botrule
\end{tabular}
\caption{Some select zero-$\PAF$ cases of order $v \neq 2 \cdot p$ for prime $p$.
The symbol \circledequal\ indicates that the two sequences of the pair are equal,
which seems to occur when $v$ is a perfect square.}
\label{tbl:Conjecture3}
\end{table}

\subsection{Periodic Golay Pairs of Length 90}\label{sec:PG90}

Inspection of Table~\ref{tbl:Conjecture3} shows a number of cases where the $d$-compressions
are of the form $([0,\dotsc,0,a],[0,\dotsc,0,b])$ where $a$ and $b$
are the rowsums of the $\PG(v)$.
Applying this pattern for $\PG(90)$ with $d\in\{18,30\}$ results in $d$-compressions
of the form $([0,0,0,0,6],[0,0,0,0,12])$ and $([0,0,6],[0,0,12])$.
For computational purposes, we chose to uncompress the pattern for $d = 18$, since 18 consists of smaller prime divisors than 30 and is easier to uncompress.
The uncompression of the pair $([0,0,0,0,6],[0,0,0,0,12])$ was performed in the three intermediate steps $18 \hookrightarrow 6 \hookrightarrow 2 \hookrightarrow 1$.
For the first step, we need only uncompress one pair by a factor of 3, so the computation time was negligible.
Filtering the results of the first step left us with 11,422 6-compressions, which we then uncompressed by a factor of 3.
On average, uncompressing each 6-compression took roughly 6 hours of computation time, so this step used roughly 8 CPU years.
Lastly, we were left with 13,267,062 2-compressions, each of which required less than one minute to uncompress.
This final uncompression step required roughly 9 CPU years in total
and produced thirty periodic Golay pairs of length 90, two of which were inequivalent.
These two inequivalent periodic Golay pairs $(A_1,B_1)$ and $(A_2,B_2)$
are given by
\begin{align*}
A_1 &= \left[{\small
\begin{array}{c@{}c@{}c@{}c@{}c@{}c@{}c@{}c@{}c@{}c@{}c@{}c@{}c@{}c@{}c@{}c@{}c@{}c@{}c@{}c@{}c@{}c@{}c@{}c@{}c@{}c@{}c@{}c@{}c@{}c@{}c@{}c@{}c@{}c@{}c@{}c@{}c@{}c@{}c@{}c@{}c@{}c@{}c@{}c@{}c}
- & - & - & - & + & + & - & + & - & + & - & - & + & + & - & + & + & - &
+ & - & - & + & + & + & + & + & - & + & + & + & - & + & + & + & - & - &
- & + & + & - & + & + & + & - & - \\ + & + & - & - & - & - & - & + & + &
+ & + & - & - & + & + & + & - & + & - & - & + & + & - & - & + & - & + &
+ & + & - & - & + & + & - & + & - & + & - & - & + & + & - & + & - & -
\end{array}}
\right], \\
B_1 &= \left[{\small
\begin{array}{c@{}c@{}c@{}c@{}c@{}c@{}c@{}c@{}c@{}c@{}c@{}c@{}c@{}c@{}c@{}c@{}c@{}c@{}c@{}c@{}c@{}c@{}c@{}c@{}c@{}c@{}c@{}c@{}c@{}c@{}c@{}c@{}c@{}c@{}c@{}c@{}c@{}c@{}c@{}c@{}c@{}c@{}c@{}c@{}c}
- & - & - & + & - & - & + & + & + & + & + & + & + & + & - & - & - & - &
+ & - & - & + & + & - & - & - & - & + & + & + & + & + & + & + & + & - &
- & + & + & - & + & - & + & - & + \\ + & - & + & - & + & + & + & + & - &
+ & - & + & + & - & + & + & + & + & - & - & + & + & + & + & - & + & - &
+ & + & + & - & + & - & - & - & - & - & + & - & - & + & - & + & - & +
\end{array}}\right], \\
A_2 &= \left[{\small
\begin{array}{c@{}c@{}c@{}c@{}c@{}c@{}c@{}c@{}c@{}c@{}c@{}c@{}c@{}c@{}c@{}c@{}c@{}c@{}c@{}c@{}c@{}c@{}c@{}c@{}c@{}c@{}c@{}c@{}c@{}c@{}c@{}c@{}c@{}c@{}c@{}c@{}c@{}c@{}c@{}c@{}c@{}c@{}c@{}c@{}c}
- & - & - & - & + & + & + & - & + & + & - & - & + & - & - & + & + & + &
+ & + & - & - & - & - & + & - & + & - & + & + & + & - & + & + & - & + &
- & - & + & - & + & - & - & - & + \\ + & + & + & + & + & + & - & + & + &
+ & - & + & - & - & - & - & - & + & + & - & - & + & - & - & + & - & + &
- & - & - & + & + & + & - & + & + & + & + & - & + & - & - & + & + & +
\end{array}}
\right],\\
B_2 &= \left[{\small
\begin{array}{c@{}c@{}c@{}c@{}c@{}c@{}c@{}c@{}c@{}c@{}c@{}c@{}c@{}c@{}c@{}c@{}c@{}c@{}c@{}c@{}c@{}c@{}c@{}c@{}c@{}c@{}c@{}c@{}c@{}c@{}c@{}c@{}c@{}c@{}c@{}c@{}c@{}c@{}c@{}c@{}c@{}c@{}c@{}c@{}c}
- & - & - & + & - & - & - & + & - & + & - & - & - & + & + & + & - & - &
+ & + & - & - & + & - & + & - & + & - & - & + & + & + & + & - & + & - &
- & + & - & + & + & + & + & - & + \\ + & + & - & - & - & - & + & - & + &
+ & + & - & - & + & + & - & + & + & + & + & + & + & + & + & + & + & + &
- & - & + & + & - & + & - & + & + & - & + & + & - & - & + & - & + & +
\end{array}}
\right].
\end{align*}
Each one of the sequences $A_1$, $B_1$, $A_2$, $B_2$ is of length 90 and is presented as 2 blocks of 45 elements each. The symbol ``$+$'' stands for $+1$ and ``$-$'' stands for $-1$.

From an incomplete perspective, it seems that $\PG(90)$ are very sparse.
The $\PG(90)$ 2-compressions never uncompressed to more than one periodic Golay pair,
while in other lengths when an uncompression yielded a solution it usually yielded many pairs.
In a previous experiment, we exhaustively classified all $\PG(90)$ 6-compressions,
of which the 11,422 pairs from our heuristic pair accounts for approximately $2\%$.
Naively extrapolating these numbers implies that there are very few $\PG(90)$,
potentially even less than length 34.

\section{Conclusion}\label{sec:conclusion}

In this paper, we develop efficient algorithms to search exhaustively for periodic Golay pairs of all lengths $v$ up to and including $v=72$. This extends significantly the range of lengths that exhaustive searches for periodic Golay pairs have previously been performed. 
Our algorithms rely primarily on two ingredients, compression and orderly generation. 
Compression is always applicable because the length of a periodic Golay pair must be an even integer, and we derived a multi-level compression method to improve cases when the length is divisible
by a small prime more than once.
Orderly generation helps to reduce the search space of our candidate compressed sequences and is also relevant for cases of complementary sequences that cannot utilize sequence compression.

We analyzed the results of our exhaustive searches and discovered patterns pertaining to new structural results on periodic Golay pairs
and proved results on the possible forms that the $(v/2)$-compression of a periodic Golay pair of order $v$ must have.
We also noticed patterns in $d$-compressions of periodic Golay pairs for values of $d$ other than $v/2$.
In particular, we noticed that in all cases we examined that
there were some values of $d$ for which there exist $d$-compressions of $\PG(v)$ with all-zero $\PAF$ vectors.
In some of these cases the $d$-compressions were of the form $([0,\dotsc,0,a],[0,\dotsc,0,b])$.
Using this observation, we used our algorithms to
uncompress $([0,0,0,0,6],[0,0,0,0,12])$
by a factor of 18 and discovered the first ever examples of periodic Golay pairs of length~$90$.

\section*{Acknowledgments}

The authors thank the reviewers for their useful comments,
including pointing out the connection between
compression and subgroup contraction.  We thank
Jonathan Jedwab for providing pointers to contraction arguments
in the literature on difference sets.

\bibliography{PG90}

\end{document}